# Role of Magnetic Anisotropy Constant Orders and Thermal Noise on Skyrmion Formation in Co/Pt Square Nanostructure

Tamali Mukherjee, V Satya Narayana Murthy[*]

Department of Physics, BITS Pilani Hyderabad Campus, Telangana, India

[*]satyam@hyderabad.bits-pilani.ac.in

**Abstract**

Skyrmions, which are topologically stable magnetic structures, have manifested promising features to be used as an information carrier in new-age, non-volatile data storage devices. In this article, Co/Pt square nano-structure with Co-free layer thickness in the range of 1 nm to 5 nm and first and second-order anisotropy constants are taken to study the controlled creation of skyrmions. The magnetization dynamics controlled by the current-induced spin transfer torque help to nucleate skyrmions by transforming the perpendicularly magnetized ground state. This process leads to a stable state of isolated skyrmions via a complex transformation of the Neel wall following its image inversion. Skyrmion numbers vary with increasing thickness as the current density gradually increases. The impact of higher-order anisotropy constants (up to the second order) on the relaxed state of a system, compared to first-order anisotropy alone, has been examined. Additionally, the effect of temperature on the formation of skyrmions has been analyzed for all thicknesses.

**Introduction**

Since 1989, various theoretical studies have shown that an asymmetric exchange interaction called Dzyaloshinskii-Moriya Interaction (DMI) [1, 2] can stabilize isolated magnetic vortices called skyrmions. They are small (a few nanometers to several microns), swirling, vortex-like structures in magnetic materials that arise due to competing effects from Heisenberg exchange interaction and DMI [3 - 6]. They are uniquely characterized by topological charge, $Q = \pm 1$ [5 - 10]. Formation and stabilization of skyrmions depend on the interplay between various energy terms, such as exchange interaction ($U_{ex}$), anisotropy ($U_{anis}$), DMI ($U_{DMI}$), Zeeman ($U_{Zeeman}$), and demagnetization energies ($U_{demag}$) [9]. Magnetic skyrmions have been observed in bulk and thin film magnetic materials with broken inversion symmetry [4, 6, 11].

In the case of thin films, the interfacial DMI between heavy metal (HM) / ferromagnetic (FM) layers such as Pt/Fe [12], Pd/Co [13, 14], Ni/Co, and Pt/CoFeB helps in stabilizing skyrmions [6, 15]. The stability of skyrmion depends on the DMI constant determined by the thickness of the heavy metal and the properties of the HM/FM interface [16, 17]. Co or CoFeB-based multilayer stacks exhibit larger DMI, making them significant in nucleation and application of skyrmions [18, 19].

Along with DMI, magnetic anisotropy energy also plays a vital role in nucleating and holding skyrmion in the system. The anisotropy constant prevents the magnetic moments from easily flipping to a random direction and secures the skyrmion in the material. The stabilization of skyrmion largely depends on the anisotropy energy as it controls the total energy landscape of the material [20]. One commonly used technology in creating and annihilating skyrmions is influenced by voltage-controlled magnetic anisotropy, as it gives good control to enhance the possibility regarding practical applicability [21, 22].

Magnetic skyrmions show promising features to be used as an information carrier for new age non-volatile memory devices due to their small size and low energy cost to manipulate them [8, 9, 23]. Easy creation and deletion are essential in the context of practical usage of skyrmions [24, 25]. Despite isolated skyrmions and lattices of skyrmions already being observed in experiments, the controlled creation still needs to be explored. In this article, we show how a current can affect the states of a perpendicularly magnetized square nano-structure by Spin Transfer Torque (STT) [26] by considering the first and second-order anisotropy constants. A stable state of isolated skyrmions is nucleated in Co/Pt multilayers by optimizing the current density. In the recent past, skyrmions have been observed at high temperatures [27, 28], and to realize the practical applicability of skyrmions, the impact of room temperature on the nucleation of skyrmions has to be well studied. This raises the question of whether this method of creating skyrmions will be effective at room temperature. In this context, the effect of temperature is examined on the final relaxed state.

**System and Simulations**

The square nano-structures we consider here are FM/HM layers of Co/Pt (Fig. 1). The free layer Co has dimensions 200 nm x 200 nm and varied thicknesses from 1 nm to 5 nm. The non-magnetic spacer layer (Pt) separates the free and ferromagnetic fixed layers. The free layer is in the x-y plane and has perpendicular magnetization along the +z direction. The fixed layer magnetization $m_p$ is along the +x direction (1,0,0).

GPU-accelerated micromagnetic software Mumax3 [29] solves the Landau-Lifshitz-Gilbert equation (LLG) using a finite difference approach. Cuboid cells of dimensions 1 nm x 1 nm x 1 nm are used. At first, the temperature is set at zero to avoid thermal fluctuations. The change in magnetization (*m*) of the free layer Co with time (*t*) is governed by the LLG equation [30 - 32],

$$\frac{\partial \bm{m}}{\partial t} = -\gamma(\bm{m} \times \bm{B}) + \alpha\left(\bm{m} \times \frac{\partial \bm{m}}{\partial t}\right) + \bm{\Gamma}_{STT} - - - - - (1)$$

Here, $\gamma$ and $\alpha$ are the gyromagnetic ratio and Gilbert damping constant, and **B** is magnetic field induction, respectively.

Spin Transfer Torque ($\bm{\Gamma}_{STT}$) is given by,

$$\bm{\Gamma}_{STT} = \gamma a\left(\bm{m} \times (\bm{m}_p \times \bm{m})\right) - \gamma b(\bm{m} \times \bm{m}_p) - - - - - (2)$$

where,

$$a = \frac{\hbar J}{2 t_{Co} e M_s} \frac{p \Lambda^2}{(\Lambda^2 + 1) + (\Lambda^2 - 1)(\bm{m}.\bm{m}_p)} - - - - - (3)$$

$$\text{and, } b = \frac{\hbar J \epsilon}{2 t_{Co} e M_s} - - - - - (4)$$

here, $\hbar$ = Planck's constant, J = current density, $t_{Co}$ = thickness of cobalt free layer, e = electronic charge, $M_s$ = saturation magnetization, p = conduction electron polarization, $\epsilon$ = coefficient of field-like torque, and for the present study, we have used p = 0.5, $\Lambda$ = 1.0, and $\epsilon$ = 0.

The material parameters essential for the simulation used are saturation magnetization ($M_s$) = 5.80 x $10^5$ A/m, exchange constant ($A_{ex}$) = 1.5 x $10^{-11}$ J/m, DMI constant ($D_{int}$) = 3 x $10^{-3}$ J/m$^2$, Gilbert damping parameter ($\alpha$) = 0.1, and anisotropy constants ($K_{u1}$ & $K_{u2}$) = 6 x $10^5$ J/m$^3$ & 1.5 x $10^5$ J/m$^3$ are taken to stabilize easy-cone state [33 - 35]. Furthermore, keeping all the other parameters same, instead of taking first and second-order magnetic anisotropy constants, we considered perpendicular magnetic anisotropy (PMA), $K_{u1}$ = 8 x $10^5$ J/m$^3$ [32, 36], to study how the anisotropy constants affect the final ground state spin configuration of the system.

Moreover, to understand the stability of skyrmions at room temperature the same method of nucleation is tested at T = 300K.

**Results and Discussions**

**Without Thermal Noise**

The skyrmion formation study uses an asymmetric current pulse of 0.5 ns positive pulse and 0.07 to 0.1 ns negative pulse (Fig. 1b) and current density (**J**) of magnitude $2 \times 10^{12} - 10 \times 10^{12}$ A/m$^2$. An easy-cone state configuration is considered by taking both first- and second-order magnetic anisotropies to see the effect of anisotropy constants. For $t_{Co}$ = 1 nm and **J** = $1 \times 10^{12}$ A/m$^2$, the system's ground state remains unchanged (free layer magnetization is along +z direction). Further, the excitation current increased to **J** = $2 \times 10^{12}$ A/m$^2$, applied in the +z direction for 0.5 ns, the spin configuration of the perpendicularly magnetized free layer changes to in-plane and forms a Neel wall (Supplementary fig. 1(b)). Once the positive current pulse is switched off, the Neel wall goes through its image inversion (Supplementary fig. 1(d)) for the same magnitude negative current density, applied for 0.1 ns. The final ground state of the system is a $180^0$ domain wall separating two reversely magnetized domains (Supplementary fig. 1(e)).

Then, we increased **J** to $2.3 \times 10^{12}$ A/m$^2$ and applied for 0.5 ns in +z direction and 0.07 ns in -z-direction. Just before the current pulse is applied (in the relaxed state), all the spins are perpendicularly aligned to the plane of the system, as shown in Fig. 2(a). The domain wall started to form (Fig. 2(b)) at t = 0.5 ns. From Fig. 2(c) (at 0.57 ns), it is evident that, by reversing the current density, the Neel wall starts transforming into its image inversion. In the absence of any applied current, a state of four isolated skyrmions of opposite core magnetization starts appearing at 0.8 ns (Fig. 2(e)) following a complicated transformation (Fig. 2(d)) at the intermediate time. Finally, from t = 3.75 ns onwards, the four skyrmions got stable at their positions along with the presence of a $180^0$ domain wall (Fig. 2(g)). The calculated diameters of the nucleated skyrmions range between 10 nm to 13 nm. Fig. 2(h), shows the topological charge (Q) variation of $t_{Co}$ = 1 nm. When the current from +z direction flipped into the -z direction, Q is found to be - 3.23 (at t = 0.55 ns). Then, after switching off the current at 0.57 ns, Q changes to 1.48 at t = 0.61 ns. Throughout the simulation time from t = 0.5 ns to t = 0.9 ns, the system undergoes complicated domain transformation triggered by the mechanism of magnetization reversal induced by STT provided by the applied current. The simulation calculates the overall Q of the sample, and the graph shows the rapid fluctuation of topological charge during the time mentioned earlier. From t = 0.9 ns onwards, the topological charge of the system becomes zero as it relaxes to a stable state of four skyrmions, with two having Q = +1 and the other two having Q = -1.

A slight variation in current density and asymmetric pulse width observes stable skyrmions. Fig. 3(a) shows that the STT provided by **J** = 2.3 x $10^{12}$ A/m$^2$ is higher than **J** = 2 x $10^{12}$ A/m$^2$. Following that in fig. 3(b), we can see the total energy ($E_{total}$), and exchange and DMI energy ($E_{Exchange + DMI}$) variation. For **J** = 2.3 x $10^{12}$ A/m$^2$, the combination of DMI energy ($E_{DMI}$) and exchange energy ($E_{Exchange}$) becomes higher and makes the total energy ($E_{total}$) also higher. As a result, the system relaxes to a stable skyrmion state along with a $180^0$ domain wall (fig. 2(c)). In the case **J** = 2 x $10^{12}$ A/m$^2$, the system relaxes to a stable state with a$180^0$ domain wall (supplementary fig. 1(e)).

With the increase in thickness, the current density required to nucleate and hold skyrmions in the nano-structure is higher. Keeping the current pulse widths same as in the case of $t_{Co}$ = 1 nm, for $t_{Co}$ = 2 nm, no skyrmion is observed. With further increasing **J** to 5 x $10^{12}$ A/m$^2$ and apply it in the +z direction for 0.5 ns and in the -z direction for 0.08 ns three skyrmions get nucleated (Fig. 4(a)). For positive pulse width (in +z direction) of 0.5 ns and negative pulse width (in -z direction) of 0.1 ns, the threshold current density to nucleate skyrmion is 7 x $10^{12}$ A/m$^2$ for 3 nm (Fig. 4(b)), 8 x $10^{12}$ A/m$^2$ for 4 nm (Fig. 4(c)), and 10 x $10^{12}$ A/m$^2$ for 5 nm (Fig. 4(d)).

Moreover, for $t_{Co}$ = 5 nm, the asymmetric pulse variation (**J** = 10 x $10^{12}$ A/m$^2$ in +z direction for 0.5 ns and 0.08 ns in -z direction) has an effect on the number of skyrmions observed. Fig. 5(a) shows three skyrmions of the same core magnetization generated in a single domain, and no domain wall is observed and in fig. 5(b) the topological charge confirms the formation of the skyrmions. It clarifies how the minute change in negative pulse width also affects the final ground state of the system. Also, this concludes that the number of nucleated skyrmions can be controlled by manipulating applied current density.

We have also carried out the micromagnetic study to compare the cone anisotropy (comes due to the first and second order constans) effect on skyrmion formation with the perpendicular magnetic anisotropy (comes due to first order constant alone) [29, 33]. From fig. 6(a), we can see that for $t_{Co}$ = 1 nm, and **J** = 2.3 x $10^{12}$ A/m$^2$ applied in +z direction for 0.5 ns and in -z direction for 0.07 ns and only considering first order anisotropy constant ($K_{u1}$), the relaxed state will be only a domain wall. However, by taking first and second order, both the anisotropy constants into consideration, the system relaxes to a stable state of four skyrmions along with a domain wall (Fig. 2(g)). Fig. 6(b) confirms that the magnetic anisotropy energy plays a crucial role in the stabilization of skyrmion, and the final relaxed spin state configuration in both cases are not the same even after applying the current of the same magnitude with identical positive and negative pulse widths.

**With Thermal Noise**

To check the nucleated skyrmion stability at room temperature (T = 300K), the threshold current pulse magnitudes kept constant as at T = 0K (**J** = 2.3 x $10^{12}$ A/m$^2$ for 1nm, 5 x $10^{12}$ A/m$^2$ for 2nm, 7 x $10^{12}$ A/m$^2$ for 3nm, 8 x $10^{12}$ A/m$^2$ for 4nm and 10 x $10^{12}$ A/m$^2$ for 5nm) and negative pulse width is varied from 0.7 to 0.1 ns. The thermal effect is introduced in the sample by adding a fluctuating thermal field term [29, 37] in the LLG equation.

$$B_{Therm} = \eta \sqrt{\frac{2\mu_0 \alpha K_B T}{B_{sat} \gamma \Delta V \Delta t}} -----(5)$$

where **η** = random vector chosen from normal distribution and changes at every step, and $B_{sat}$ = saturation magnetic field. The other constants are, α = Gilbert damping parameter, $K_B$ = Boltzman constant, T = temperature applied, ɣ = gyromagnetic ratio, ΔV = unit cell volume and, Δt = time step. The sixth order Runge-Kutta-Fehlberg solver is used to solve the LLG equation [38].

For $t_{Co}$ = 1nm, at T = 300K, when **J** = 2.3 x $10^{12}$ A/m$^2$ applied in +z direction for 0.5ns and in -z direction for 0.07ns, no skyrmions formed. Only one skyrmion was observed when negative pulse width has been increased to 0.09ns. For $t_{Co}$ = 2 and 3nm, the maxium no. of skyrmions is 2, which is lesser than observed at 0K. It is to be noted that, temperature assisted in forming skyrmions at higher thicknesses of free layer Co [39]. The maximum number of skyrmions observed at T = 300K is 3 at $t_{Co}$ = 4nm (supplementary fig. 2(g)) for **J** = 8 x $10^{12}$ A/m$^2$ applied in +z direction for 0.5ns and in -z direction for 0.09ns, while at T = 0K, with the same current magnitude and pulsewidth, no skyrmions were formed. Furthermore, increasing the Co layer thickness to 5nm, the maximum number of skyrmions obtained by varying the pulsewidth was 3. The diameters of the simulated skyrmions vary from 14nm to 22nm. Although how the combination of strength of negative pulse width and temperature exactly plays a role in varying the number of skyrmions generated in the relaxed state could not be precisely understood, further studies are required to better comprehend this phenomenon.

**Conclusions**

The controlled creation of skyrmions with nano-second asymmetric current pulses in a square Co/Pt nano-structure, with Co thicknesses range of 1 – 5 nm, is studied by employing micromagnetic simulation via Mumax3. A maximum number of 4 skyrmions (at T = 0K) is observed, for current densities ranging between $2 \times 10^{12}$ A/m$^2$ to $10 \times 10^{12}$ A/m$^2$ applied in +z direction for 0.5 ns and in -z direction for 0.07 ns to 0.1 ns (modulated in steps of 0.005 ns). To understand the role of magnetic anisotropy in stabilizing skyrmions, cone anisotropy (first-order + second-order anisotropy constants) and first-order constant alone are used in the study. A slight change in the negative pulse width causes a significant change in the relaxed state through the complicated transformation of the spin system. The effect of thermal noise on the skyrmion formation have also been studied. Simulating skyrmions at room temperature (T = 300K) is more feasible when the free layer thickness is high ($t_{Co}$ = 4nm and 5nm).

**Acknowledgments**

We acknowledge BITS Pilani Hyderabad Campus for providing the Sharanga High-Performance Computational facility to carry the above work.

**Figures**

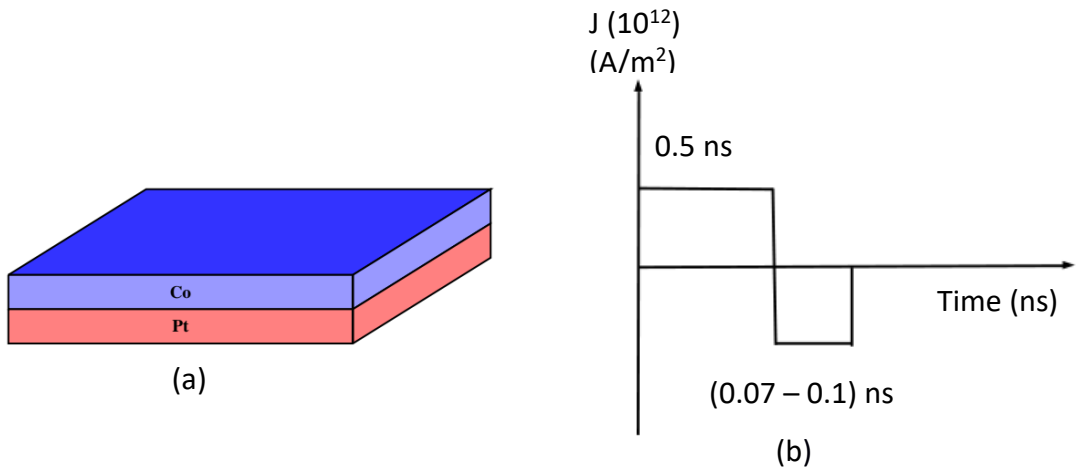

**Fig. 1:** (a) 200 nm x 200 nm Co/Pt square nano-structure with Co free layer (thickness varied from 1 nm – 5 nm), and (b) Nano-second Current pulse applied to Co/Pt. The layers are in the xy - plane, and the current flows in the +z and -z directions.

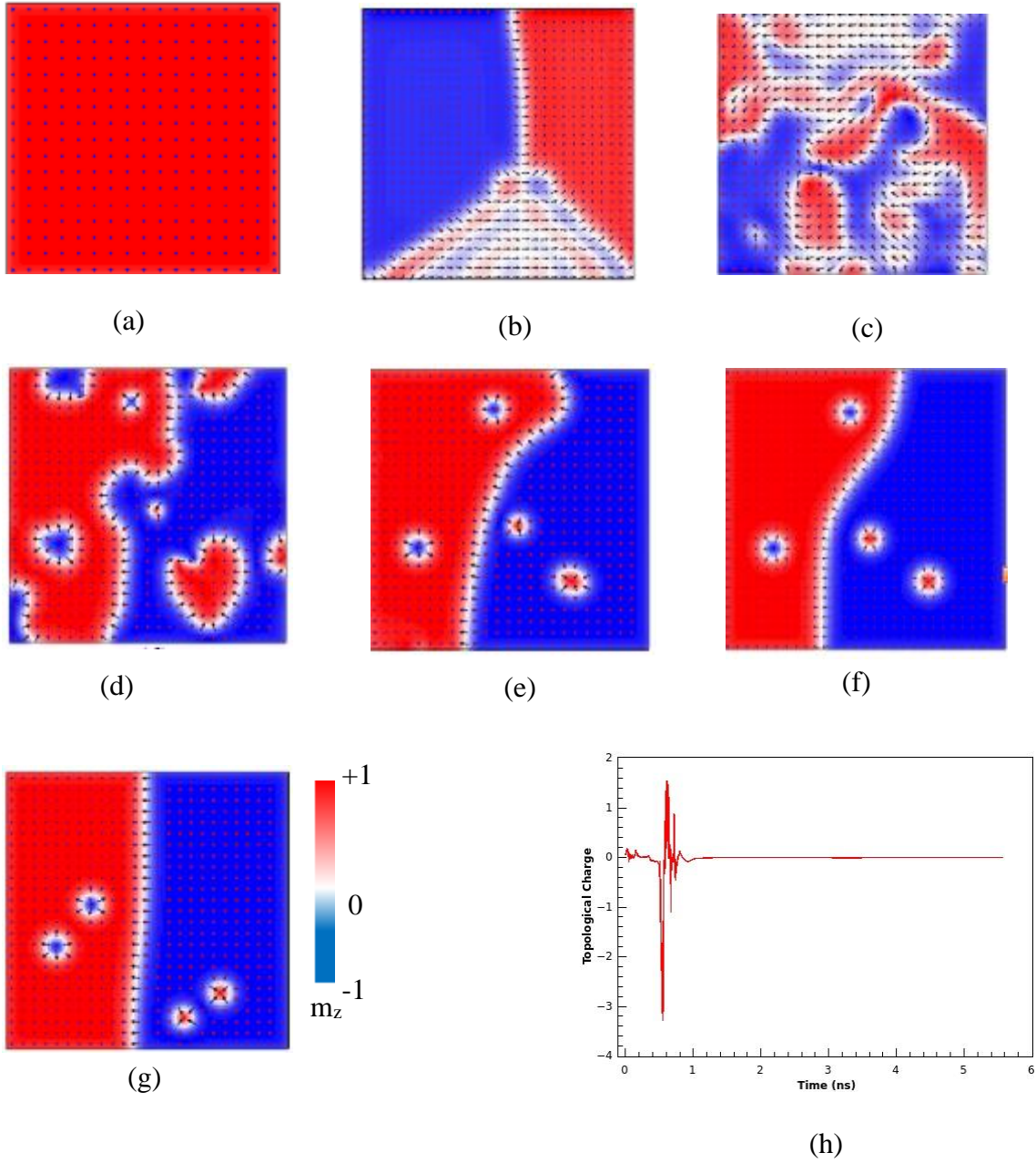

**Fig. 2:** Spin configurations for $t_{Co}$ = 1 nm with applied current density J = 2.3 x $10^{12}$ A/$m^2$ in +z direction for 0.5 ns and in -z direction for 0.07 ns at (a) t = 0 ns, (b) t = 0.5 ns, (c) t = 0.57 ns, (d) t = 0.65 ns, (e) t = 0.80 ns, (f) t = 1 ns, and (g) t = 3.75 ns. The arrows are projection of **m** onto xy plane. The colour bar denoting $m_z$ is shown. (h) Topological charge of the sample vs. smulation time. The constant topological charge = 0 denotes four stable skyrmions, with each two having opposite core magnetization +1 and -1, respectively.

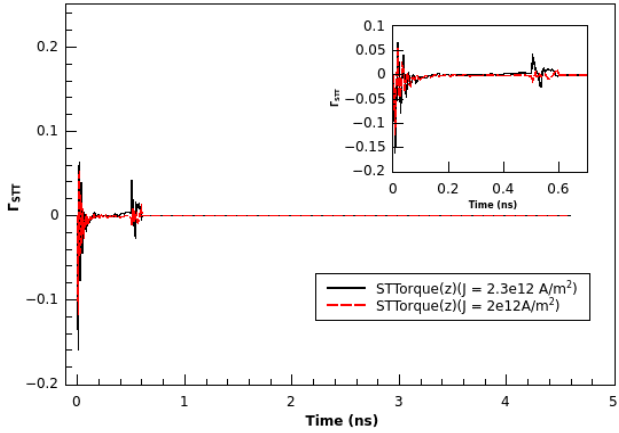
(a)

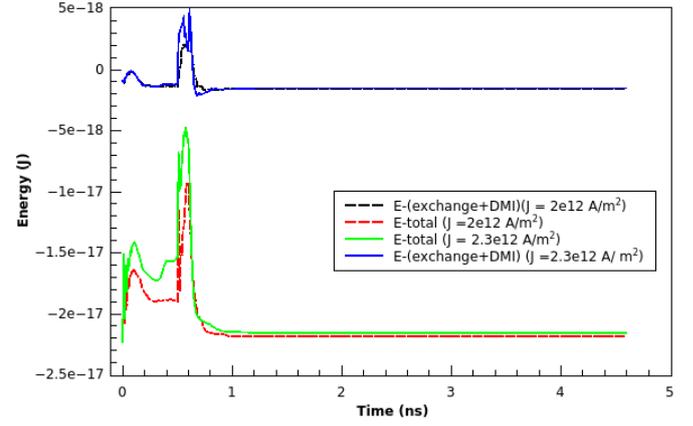
(b)

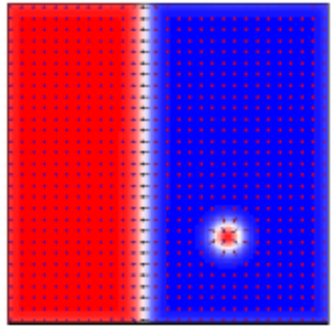
(c)

**Fig. 3:** For a Co free layer thickness of 1 nm (a) Variation of the z component of the STT with simulation time (the inset figure explains the spin transfer torque experienced for the initial 0.5 ns when the current pulse is on), (b) Energy landscape of the sample for J = 2 x $10^{12}$ A/m$^2$ (dashed lines) and 2.3 x $10^{12}$ A/m$^2$ (solid lines) in +z direction for 0.5 ns and in -z direction for 0.1 ns, and (c) Final relaxed state of the system for J = 2.3 x $10^{12}$ A/m$^2$. The spin transfer torque provided by this specific current density and pulse width relaxes one skyrmion having Q = +1. The colour bar is the same as in fig. 2.

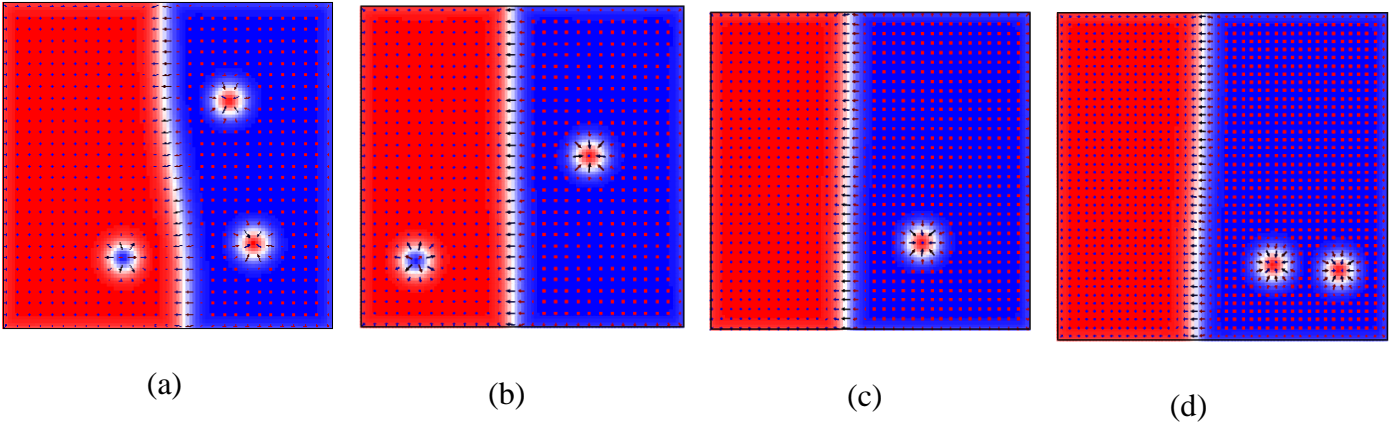

**Fig. 4:** Stable state of isolated skyrmions for Co layer thickness (a) 2 nm ($J = 5 \times 10^{12}$ A/m$^2$), (b) 3 nm ($J = 7 \times 10^{12}$ A/m$^2$), (c) 4 nm ($J = 8 \times 10^{12}$ A/m$^2$), and (d) 5 nm ($J = 10 \times 10^{12}$ A/m$^2$). The current density is applied in +z direction for 0.5 ns and in the -z direction for 0.1 ns. All the snapshots are taken after 10 ns. The color bar is the same as in fig. 2.

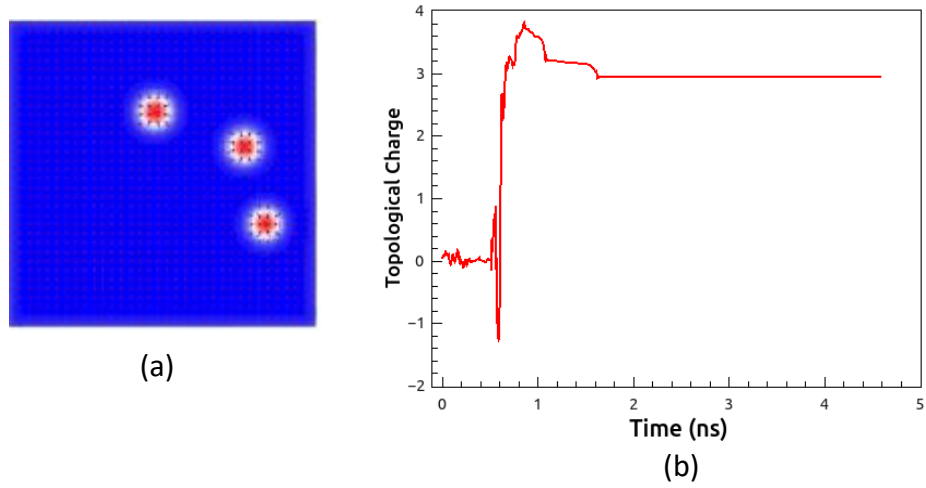

**Fig. 5:** (a) A stable state of 3 skyrmions is observed for $t_{Co}$ = 5 nm when J = 10 x $10^{12}$ A/$m^2$ is applied for 0.5 ns in +z direction and for 0.08 ns for in -z direction, and (b) Topological charge vs. simulation time plot, confirms the relaxed state is holding 3 skyrmions of same core magnetization, making the total topological charge Q = +3. The color bar is the same as in fig. 2.

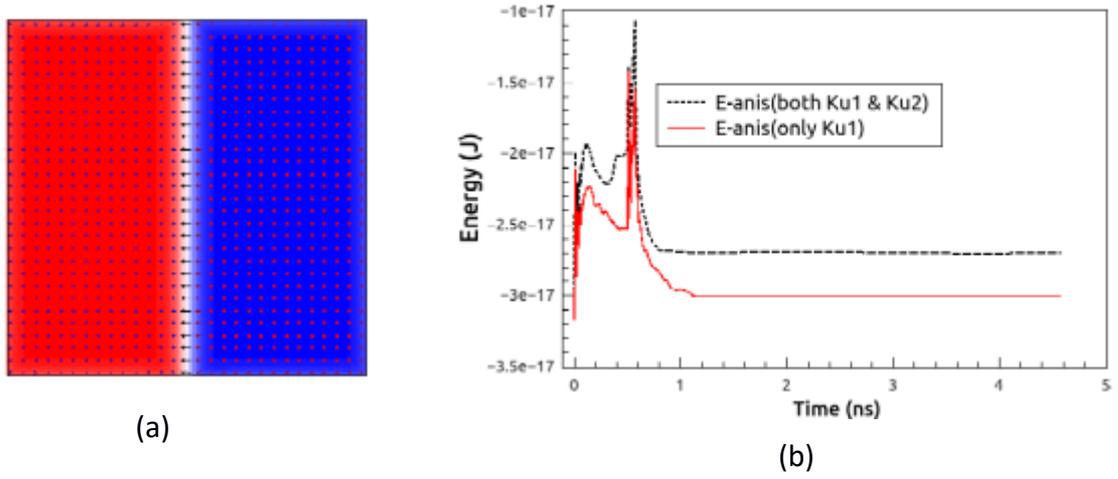

**Fig. 6:** Effect of only first order anisotropy constant on skyrmion formation: (a) Final ground state achieved with PMA ($K_{u1}$= 8 x $10^5$ J/m$^3$) for $t_{Co}$= 1nm and J = 2.3 x $10^{12}$ A/m$^2$ in +z direction for 0.5 ns and in -z direction for 0.07 ns, no skyrmion is observed and (b) Anisotropy energy comparison plot with